# Insights from a computational analysis of the SARS-CoV-2 Omicron variant: Host-pathogen interaction, pathogenicity, and possible therapeutics

Md Sorwer Alam Parvez[1,2], Manash Kumar Saha[2], Md. Ibrahim[2], Yusha Araf[2], Md. Taufiqul Islam[2], Gen Ohtsuki[1,*], Mohammad Jakir Hosen[2,*]

[1]Department of Drug Discovery Medicine, Kyoto University Graduate School of Medicine, 53 Shogoin-Kawahara-cho, Sakyo-ku, Kyoto 606-8507, Japan

[2] Department of Genetic Engineering & Biotechnology, Shahjalal University of Science & Technology, Sylhet-3114, Bangladesh

*Correspondence:
Prof. Mohammad Jakir Hosen: jakir-gen@sust.edu

Prof. Gen Ohtsuki: ohtsuki.gen.7w@kyoto-u.ac.jp

**Abstract**

Prominently accountable for the upsurge of COVID-19 cases as the world attempts to recover from the previous two waves, Omicron has further threatened the conventional therapeutic approaches. Omicron is the fifth variant of concern (VOC), which comprises more than 10 mutations in the receptor-binding domain (RBD) of the spike protein. However, the lack of extensive research regarding Omicron has raised the need to establish correlations to understand this variant by structural comparisons. Here, we evaluate, correlate, and compare its genomic sequences through an immunoinformatic approach with wild and mutant RBD forms of the spike protein to understand its epidemiological characteristics and responses towards existing drugs for better patient management. Our computational analyses provided insights into infectious and pathogenic trails of the Omicron variant. In addition, while the analysis represented South Africa's Omicron variant being similar to the highly-infectious B.1.620 variant, mutations within the prominent proteins are hypothesized to alter its pathogenicity. Moreover, docking evaluations revealed significant differences in binding affinity with human receptors, ACE2 and NRP1. Owing to its characteristics of rendering existing treatments ineffective, we evaluated the drug efficacy against their target protein encoded in the Omicron through molecular docking approach. Most of the tested drugs were proven to be effective. Nirmatrelvir (Paxlovid), MPro 13b, and Lopinavir displayed increased effectiveness and efficacy, while Ivermectin showed the best result against Omicron.

**Introduction**

COVID-19 pandemic by the SARS-CoV-2 (also known as coronavirus) has pulverized the health care system of the world since November 2019, which changed our lives and caused strict measures to prevent the spread of infection (WHO, 2021, October 13). Currently, most countries are threatened by the 3rd to 6th wave of this severe acute respiratory disease, and the entire world is trying to combat it (Hiscott et al., 2020). It's a situation constantly changing and evolving (Anwar et al., 2020). Multiple forms of this virus, including alpha, beta, gamma, and delta have shown its rampage, and most recently, the omicron form is circulating over the world with a hot spot of more than 30 mutations in the spike protein (Karim et al., 2021; WHO, 2021, November 26).

Omicron, a newly evolved and very highly-infectious coronavirus variant (B.1.1.529), was designated as a variant of serious concern by the World Health Organization on November 26, 2021 (Collie et al., 2021). Since the first case report in Botswana on November 11, 2021, Omicron has spread to 108 countries and infected 150-thousand patients within a month, despite greater surveillance. While it is too early to assess exact severity, preliminary findings suggest that Omicron has a less clinical presentation and 4.9% lower hospital admission rates (Jassat et al., 2021). It is the most highly-altered version, similar to those reported in earlier variants of concern, linked to its increased transmissibility and partial resistance to vaccine-induced immunity (Collie et al., 2021; Torjesen, 2021; Gu et al., 2021). Omicron was born into a COVID-19-weary world and repleted with further anxiety and distrust at the pandemic's extensive detrimental social, emotional, and economic consequences (Karim and Karim 2021)

The laboratories chasing the Omicron variant have yet thoroughly defined its epidemiologic characteristics. The features of DNA sequence alone cannot be used to determine them, which causes a diagnostic challenge. Concerning the spike protein of the Omicron variant, mutations were reported in the S protein. The alterations in the S protein receptor-binding domain (RBD) may influence its infectivity and antibody resistance, as RBD is necessary for binding with host angiotensin-converting enzyme 2 (ACE2) during the early infection process. The binding free energy (BFE) between the S RBD and the ACE2 has been demonstrated proportional to viral infectivity in several investigations. Moreover, mutations in the nucleocapsid protein have also been reported in Omicron, which helps viral proliferation (Gu et al., 2021). Increased transmissibility, better viral binding affinity, and higher antibody escape would have all been linked to these mutations (Karim et al., 2021).

The Omicron variant has currently become a great concern for the world. Basic research is required to unveil its molecular consequences via gene mutations, which resulted in changes in infectivity, pathogenicity, and antigenic escape potential (Harvey et al., 2021). Both occurrence area and the variant of Omicron origin are also unclear. To face the challenge of Omicron, it is urgent to test possible therapeutics and the effectiveness of available vaccines. Therefore, our study aimed to elucidate the novelty of Omicron from other variants: molecular mechanism of its high infectious ability and less pathogenicity. We also analyzed the effectiveness of current promising drugs against this variant. Our findings should provide novel insights on the structural and functional impact of mutations in Omicron, the impact during host interaction, and possible therapeutics for combatting this highly infectious variant.

## Materials and Methods

### Retrieval of the Sequences:

The complete genome sequences of all notable SARS-CoV-2 variants including the variant of concern (VOC) and the variant of interest (VOI) were retrieved from the GISAID database (www.gisaid.org). We collected a total number of 30 variants along with South Africa variant B.1.1.529 (Omicron) from this database. Additionally, the sequence of the Wuhan SARS-CoV-2 was also retrieved and considered as the reference for the comparative analysis.

### Multiple Sequence Alignment and Phylogenetic Tree Reconstruction:

We performed multiple sequence alignment (MSA) using MUSCLE alignment tools (Edgar, 2004). Further, we used this MSA file for the reconstruction of a phylogenetic tree. IQ-TREE was adopted for the reconstruction of the tree with Maximum Likelihood (ML) method (Nguyen et al., 2015). To identify the best-fit substitution model, we used ModelFinder for the model test (279 models) and selected the best-fit substitution model (GTR+F+R2) based on Bayesian Information Criterion (BIC) which is a criterion for model selection and generally model with lower BIC is

considered as good (Kalyaanamoorthy et al., 2017). BIC is calculated with the following equations:

$$BIC = k\ln(N) - 2\ln(L)$$

Here,
k= number of parameters estimated by the model
N= the number of data points in the number of observations or equivalently, the sample size
L= the maximized value of the likelihood function of the model

Besides, we performed both SH-aLRT and ultrafast bootstrap to assess branch supports where both were set to 1000. UFBoot2 was used for this bootstraps assessment operation (Hoang et al., 2018). Finally, we employed an iTOL online tool for the visualization and analysis of the reconstructed phylogenetic tree (Letunic and Bork, 2021).

**Identification of the Nucleotide Variations:**

The MSA file was analyzed by MEGAX software to identify the nucleotide variations in all variants, considering the Wuhan strain as a reference (Kumar et al., 2018).

**Prediction of the Encoded Proteins and Identification of the Variations:**

To predict the genes and their encoded proteins in the variant genome, we used FGENESV (uses pattern recognition and Markov chain models) of Softberry (http://linux1.softberry.com/berry.phtml) viral gene prediction tools. The predicted genes and proteins were further confirmed using the Basic Local Alignment Search Tool (BLAST) of NCBI (https://blast.ncbi.nlm.nih.gov/Blast.cgi). A pair-wise alignment of each protein with its corresponding protein of the reference strain was further performed to identify the amino acid variations in each protein by adopting Clustal omega, and visualized and analyzed in MVIEW (Sievers and Higgins, 2014; Brown et al., 1998).

**Modeling of the Mutant RBD of Spike Protein and Validation:**

The 3D crystal structure of wild RBD of spike protein was retrieved from the Protein Data Bank database (https://www.rcsb.org/) with the accession number of PDB ID: 6M17 (Berman et al., 2000). This structure was cleaned by removing water, ligand, and other complexed molecules using the PyMOL (DeLano, 2002). We modeled the 3D structure of mutant RBD of Omicron variant using SWISS-MODEL (https://swissmodel.expasy.org/) web server (Kiefer et al., 2009). Then, the generated 3D structure was validated by using ERRAT and PROCHECK (Colovos and Yeates, 1993; Laskowski et al., 1993).

**Molecular Docking of RBD of Spike Protein with Human Receptors:**

We analyzed the interaction of both wild and mutant RBD of spike protein with human receptors ACE2, NRP-1, BSG, and DPP4 through protein-protein molecular docking using FRODOCK tools (Ramírez-Aportela et al., 2016). We selected these four human receptors, as previous studies demonstrated that spike protein interacts with them. The 3D structure of the receptor proteins was retrieved from the PDB except for NRP-1. Due to the unavailability of NRP-1 3D structure in PDB, we retrieved it from the Alphafold database (Jumper et al., 2021). Then, we prepared the input PDB file by converting it to PQR format using PDB2PQR, which is a Python-based structural conversion utility (Dolinsky et al., 2004). We used CHARMM force field during the protein-protein docking simulation (Vanommeslaeghe et al., 2010). Finally, we calculated and obtained the binding energy of the binding interaction by using the PISA server (https://www.ebi.ac.uk/msd-srv/prot_int/cgi-bin/piserver).

**Analysis of the Effectiveness of Promising Drugs:**

We analyzed ten promising drugs targeting the main (3CL) protease protein of SARS-CoV-2 including Nirmatrelvir, Ritonvir, Ivermectin, Lopinavir, Boceprevir, MPro 13b, MPro N3, GC-373, GC376, and PF-00835231 (Jin et al., 2020; Zhang et al., 2020; Vandyck and Deval, 2021). The PDB structures of these drugs were retrieved from the DrugBank (https://go.drugbank.com/) and PubChem (https://pubchem.ncbi.nlm.nih.gov/) (Wishart et al., 2018; Kim et al., 2016). The 3D structure of the wild main protease was retrieved from PDB ID: 6WTK. We modeled the mutant by SWISS-MODEL (Kiefer et al., 2009). Prior to the molecular docking, we removed water molecules, ligands, and other complex molecules from the 3D structures. Polar hydrogen atoms and required charges for the energy minimization were further added. Molecular docking was performed by using AutoDock Vina tools (Trott and Olson, 2010). We set parameters of the grid box to size 40 Å × 64 Å × 64 Å (x, y, and z) and center -16.773 × -15.229 × 13.709 (x, y, and z). Further, we used PyMOL for the analysis and visualization of the protein-ligand complex molecules (DeLano, 2002). The 2D diagrams of Protein-ligand interaction were generated with Discovery Studio (Biovia, 2017).

## Results

### Phylogenetic Analysis:

First, we reconstructed a phylogenetic tree from the multiple sequence alignment of all notable variants, including South Africa B.1.1.529 (Omicron). We further reconstructed an un-rooted phylogenetic tree through maximum likelihood (ML) methods to find closely related variants. Surprisingly, we found that the Omicron variant was very closely related to SARS-CoV-2 variants of Germany B.1.620 (Fig 1). Interestingly, the Switzerland B.1.1.318 variant was also localized close to the Omicron variant. All variants were centered at the Wuhan strain (Fig 1, red circle) which was the very early strain of SARS-CoV-2.

### Identification of the Nucleotide Variations

Variations in the genome sequences of all notable variants from the alignment file were identified by comparing them with the Wuhan strain. We found that the variant South Africa B.1.1.529 (Omicron), UK B.1.1.7+S494P, and Russia AT.1 were highly mutated (Table 1). However, most of the mutations in the UK B.1.1.7+S494P and Russia AT.1 variant were synonymous, but Omicron contained the maximum number of non-synonymous mutations (50 non-synonymous mutations). Remarkably, we found most of these non-synonymous mutations to locate in the spike protein sequence. Importantly, some of the deletion and insertion was found in the consecutive bases of Omicron affect encoded proteins (Table 2).

### Identification of the Mutations in the Proteins

FGENESV and further pairwise alignment analysis demonstrated that the Omicron variant had mutations across polyprotein ab, spike protein, envelope protein, membrane glycoprotein, and nucleocapsid phosphoprotein (Table 3). Most numbers of the mutations were found located in spike protein. In the case of polyprotein ab, mutations occurred in papain-like protease nsp3, nsp4, 3C like protease nsp5, nsp6, RNA dependent RNA polymerase nsp12, and proofreading exoribonuclease nsp14. Interestingly, deletions of three consecutive amino acids at positions 31 to 33 were detected in nucleocapsid phosphoprotein.

### Modeling of the Mutant RBD of Spike Protein and Validation:

The 3D modeling of mutant RBD of spike protein was done by SWISS-MODEL homology modeling using wild-type RBD of spike protein (with PDB ID 6M17) as a template. We further validated the obtained 3D model through ERRAT and PROCHECK tools. ERRAT validation revealed the overall quality factor of this model was 97.5 (Fig 2). Ramachandran plot analysis by PROCHECK revealed that 91.8% of its residues were in the most favoured regions, and 7.6% were in additional allowed regions (Fig 2). A model would be considered as a good quality and high-reliability if it has over 90% of its residues in the most favoured regions. All these validation scores suggested that this model was highly reliable to use for further analysis.

**Molecular docking of RBD of Spike Protein with Human Receptors:**

Next, we performed molecular docking analysis to investigate the impact of mutations in the RBD spike protein in interaction with human receptors. We used four previously reported receptors, including ACE2, NRP-1, DPP4, and BSG in this analysis (Scialo et al., 2020; Daly et al., 2020; Yang et al., 2021; Essahib et al., 2020). Docking analysis revealed that the binding energy for interaction with ACE2 was decreased from -15.9 to -17.2 while increased for NRP1 from -27.2 to -22.9, compared to wild-type RBD (Table 4 & Fig 3). The other two receptors showed higher binding energy in both wild and mutant type RBD, compared to ACE2 and NRP1.

**Effectiveness of the Promising Drugs:**

Nirmatrelvir, Ritonvir, Ivermectin, Lopinavir, Boceprevir, MPro 13b, MPro N3, GC-373, GC376, and PF-00835231 were reported as effective against SARS-CoV-2, and most of them were currently in the clinical trials. To investigate the effectiveness of these drugs, we performed molecular docking against the main protease protein of Omicron. We found that mutations in the main protease that occurred in the Omicron variant didn't significantly affect the binding energy for the interaction between these drugs and the main protease (Table 5). The binding affinity increased for Nirmatrelvir, MPro 13b, and Lopinavir (Fig 4), whereas no changes for Ivermectin, MPro N3, and GC-373. The lowest binding energy was found for Ivermectin against both wild and mutant main protease. The binding site for all drugs was similar, however, the interacted amino acids were different (Fig 5).

**Discussion**

Twenty-three months following the first emerging cases of COVID-19 alongside its several variant classifications, another VOC, known as Omicron or B.1.1.529, was reported on November 26, 2021. While the world attempts to overcome the repercussions of COVID-19, the high transmissibility and pathogenesis of the variants act as reversals. The conventional nature of RNA viruses to cause mutations within their genome raised the concerns associated with transmission and infection degrees. The following example would include the Delta variant, which has claimed millions of lives all around the world (Kannan et al., 2021). On the other hand, the recently emerging the Omicron variant is the fifth VOC after Alpha, Beta, Gamma, and Delta. However, while the variants emerged through mutations, the mutational profile of Omicron is significantly different in comparison to the other variants; even though some genomic alterations resemble that of Beta and Delta, it is not exactly similar at the molecular level (Poudel et al., 2022). The spike protein or S protein of Omicron is known to be the major site of mutation, which is also labeled to have increased infectivity and transmissibility owing to its protein-specific mutations. A similar mutation trend followed by previous VOCs, alongside other changes within the viral genome, also raises concerns associated with antiviral drug effectiveness, antibody therapies, and vaccine-conferred immunity (Kannan et al., 2022; Hoffmann et al., 2021). Upholding these concerns, the Omicron variant has reportedly not only rendered vaccines less effective but has also contributed to compromising antibody-based therapies, which have been proving to be the conventional lifesavers. Consequently, the need for novel effective antivirals and an evaluation of their targeted action against the virus remains of crucial significance. Through comprehensive evaluations, our study identified the structure-based indifferences of the now emerging and dominant variant, Omicron. Further, this study elucidated not only the associated interactions between the RBD of spike protein and human receptors but also the effectiveness of the existing antiviral drugs.

Our extensive evaluations into the phylogenetic tree analysis initially indicated that the South Africa B.1.1.529 variant (i.e., Omicron) was quite similar to the SARS-CoV-2 B.1.620 variant of Germany. Surprisingly, B.1.620 was prevalent in Africa before emerging the Omicron variant which may support to conclude B.1.620 as the origin of the Omicron variant (Dudas et al., 2021). Additionally, The B.1.620 variant had the D614G mutation, which was responsible for an increased SARS-CoV-2 infection pattern (Yurkovetskiy et al., 2020). Therefore, it can be credited that the Omicron variant follows a similar infectivity trend owing to its phylogenetic similarity that contributes to the current surging COVID-19 cases worldwide. Besides, the South Africa B.1.1.529 Omicron variant was found to have an overall of 50 non-synonymous mutations - a majority of which were found in the spike protein. Moreover, the identification-based analysis revealed that the Omicron variant consisted of mutations in polyprotein ab, spike protein, envelope protein, membrane glycoprotein, and nucleocapsid phosphoprotein. And polyprotein ab is cleaved into several non-structural proteins. We also found mutations in these proteins, including papain-like protease nsp3, nsp4, 3CL protease nsp5, nsp6, RNA dependent RNA polymerase nsp12, and proofreading exoribonuclease nsp14. Seemingly, a study by Zhu et al. (2021) observed through Drosophila viability assays that nsp6 was one of the most pathogenic SARS-CoV-2 genes, capable of triggering lethal consequences individually and, at the same time, was labeled as one of the primary determinants of COVID-19 pathogenesis (Zhu et al., 2021). Mutations in this protein could affect the intracellular survival of the virus and could also make a significant modification in viral pathogenicity (Benvenuto et al., 2020). Three consecutive deletions and a substitution mutation in the genome sequence of Omicron possibly indicate a reduced pathogenicity. Mutations of nsp3, a major protein for the SARS-CoV-2 replication, suggest a lower replication rate and infectivity. These two proteins, along with nsp4 and nsp5, are known for double-membraned vesicle (DMV) inductions and localizations of cleaved maps (Gorkhali et al., 2021; Angelini et al., 2013). While nsp12 and nsp14, which are required to mediate polymerase and exonuclease activities, had also been shown genetic alterations presumably affecting their viral load, other structural proteins were also mutated. These include the Spike (S) protein which is responsible for facilitating the membrane fusion and viral entry (Tortorici and Veesler, 2019); the Envelope (E) protein which is contributory to virus morphogenesis and pathogenesis (Hogue and Machamer, 2014); the Membrane (M) protein which aids membrane fusion through its initial attachment to the S protein and surface receptors of the host (Lai and Cavanagh, 1997; Fleming et al., 1989); the Nucleocapsid (N) protein moderates replication and viral RNA synthesis, transcription and metabolism associated with infected cells and additionally provides stability to the RNA inside the cell (Cong et al., 2019; Huang et al., 2004; Nelson et al., 2000; Stohlman et al., 1988). While these generalized protein roles and their mutations may help hypothesize Omicron as less pathogenic than others, only further research into their gene-specific mutations of the Omicron variant may act as better pointers for characteristic identification.

With a maximum of mutations in the spike (S) protein, the mutant RBD of that very protein was modeled and validated for further analysis. We studied the chosen human receptors, including ACE2, NRP1, DPP4, and BSG, through molecular docking processes to understand their interactions with the mutant RBD of the spike protein. Notably, the results indicated a decreased binding energy for interacting mutant RBD with ACE2 and a significantly increased binding energy for interacting mutant RBD with NRP1, compared to the wild-type RBD. However, both DPP4 and BSG showed higher binding energies in either form, compared to the former two human receptors. Spike protein RBD of the Omicron variant contains 11 mutations, and they may be the responsible elements for increasing binding affinity for the interaction with ACE2. Mutations in RBD occurred for an optimization of the binding affinity, which would be advantageous for the virus to enhance its transmissibility. Barton et al. reported that three mutations of RBD: N501Y, E484K, and S477N enhance the binding affinity for interaction with ACE2 (Barton et al., 2021). Surprisingly, all these mutations were available in the spike protein RBD of the Omicron variant, and they may be responsible for increased binding affinity. Acknowledging the characteristic label of infecting 70 times faster than the deadly Delta variant and the initial COVID-19 strain but likely being less severe (Luo et al., 2021), a generalized hypothesis may be provided alongside the results of our study. Owing to its greater infectivity but lower pathogenicity, as a comparison to its receptor binding capacity, it may be hypothesized that ACE2 is responsible for increased infectivity whereas NRP1 is associated with increased pathogenicity. Therefore, in cases of the Omicron variant, the increased binding affinity for ACE2 corresponds to its greater infection rate whereas, decreased binding affinity for NRP1 may correspond to a decreased pathogenicity.

Furthermore, an analysis based on the drug effectivity including Nirmatrelvir, Ritonvir, Ivermectin, Lopinavir, Boceprevir, MPro 13b, MPro N3, GC-373, GC376, and PF-00835231, was also conducted. The evaluation of these drugs to determine their interaction with their targeted main protease of the Omicron variant revealed that mutations within the major interacting protein did not hamper much the binding energy at all, except for Boceprevir and GC-376, which showed increased binding energy. The increased binding affinity of Nirmatrelvir (Paxlovid), MPro 13b, and Lopinavir may indicate their greater drug efficacy against this Omicron variant compared to previous variants. This result is also supported by the recent announcement by Pfizer about their drug Paxlovid being effective against the Omicron variant (Reuters, 2021, December 23). However, Ivermectin showed the highest binding affinity and may be the most effective drug candidate against the Omicron variant. While these hypotheses hold great value and may provide significant insights into the therapeutic strategies, further research is crucial to authenticate these statements.

## Conclusion

The world is now afraid of the highly infectious omicron variant, and research is required to know about this variant. Our study gave an insight into its probable molecular consequences about infectivity and pathogenicity of the Omicron variant. The study also demonstrated that the highly infectious B.1.620 strain may be the origin of the Omicron variant, and mutations in all major proteins made omicron less pathogenic. Through docking analysis, we revealed that the mutations in spike protein increased its binding affinity for its main receptor ACE2 while decreased binding affinity for its co-receptor NRP-1. All the promising drugs that target the main protease would also be effective against this variant; however, Ivermectin shows the strongest binding affinity, and Nirmatrelvir (Paxlovid), MPro 13b, and Lopinavir may be more effective against this variant.

## Funding

This work was supported by grants from the Mitsubishi Foundation, the Takeda Science Foundation (to G.O.).

## Conflict of interest

The authors have declared no competing interest.

## Credit Author Statement

**Md Sorwer Alam Parvez:** Conceptualization, Methodology, Formal analysis, Data Interpretation, Validation, Visualization & Original draft preparation; **Manash Kumar Saha**: Methodology, Software, Visualization; **Md Ibrahim:** Formal Analysis; **Yusha Araf:** Formal analysis, Original draft preparation & editing; **Md Taufiqul Islam:** Validation; **Gen Ohtsuki:** Supervision, Writing-review & editing; **Mohammad Jakir Hosen**: Supervision, Writing-review & editing

Anwar, S., Nasrullah, M., and Hosen, M.J. (2020). COVID-19 and Bangladesh: challenges and how to address them. *Frontiers in Public Health*, 8, 154. doi: 10.3389/fpubh.2020.00154

Barton, M.I., MacGowan, S.A., Kutuzov, M.A., Dushek, O., Barton, G.J. and van der Merwe, P.A., 2021. Effects of common mutations in the SARS-CoV-2 Spike RBD and its ligand, the human ACE2 receptor on binding affinity and kinetics. *Elife*, *10*, p.e70658

Benvenuto, D., Angeletti, S., Giovanetti, M., Bianchi, M., Pascarella, S., Cauda, R., ... & Cassone, A. (2020). Evolutionary analysis of SARS-CoV-2: how mutation of Non-Structural Protein 6 (NSP6) could affect viral autophagy. *Journal of Infection*, *81*(1), e24-e27. doi: 10.1016/j.jinf.2020.03.058

Berman H.M., Westbrook J., Feng Z., Gilliland G., Bhat T.N., Weissig H., Shindyalov I.N., Bourne P.E. (2000). The protein data bank. *Nucleic Acids Research*, 28(1):235-242. doi: 10.1093/nar/28.1.235

Biovia, D. S. (2017). Discovery studio modeling environment.

Brown, N.P., Leroy, C., & Sander, C. (1998). MView: a web-compatible database search or multiple alignment viewer. *Bioinformatics (Oxford, England)*, *14*(4), 380-381. doi:10.1093/bioinformatics/14.4.380

Collie, S., Champion, J., Moultrie, H., Bekker, L.G., and Gray, G., 2021. Effectiveness of BNT162b2 Vaccine against Omicron Variant in South Africa. *New England Journal of Medicine*. doi: 10.1056/NEJMc2119270

Colovos, C., & Yeates, T.O. (1993). Verification of protein structures: patterns of nonbonded atomic interactions. *Protein Science*, 2(9), 1511-1519. doi: 10.1002/pro.5560020916

Cong, Y., Ulasli, M., Schepers, H., Mauthe, M., V'kovski, P., Kriegenburg, F., ... & Reggiori, F. (2020). Nucleocapsid protein recruitment to replication-transcription complexes plays a crucial role in coronaviral life cycle. *Journal of virology*, *94*(4), e01925-19. doi:10.1128/jvi.01925-19

Daly, J.L., Simonetti, B., Klein, K., Chen, K.E., Williamson, M.K., Antón-Plágaro, C., Shoemark, D.K., Simón-Gracia, L., Bauer, M., Hollandi, R. and Greber, U.F., 2020. Neuropilin-

**Figure Legends**

**Fig 1:** Maximum likelihood unrooted phylogenetic tree of all notable SARS-CoV-2 variants. South Africa B.1.1.529 (Omicron) was very close to Germany B.1.620. All the variants were centered with the Wuhan strain (red circle). A scale indicates genetic variation, defined as the number of substitutions per nucleotide site.

**Fig 2:** Validation of the 3D model. **A,** ERRAT validation assessment. **B,** Ramachandran Plot by PROCHECK.

**Fig 3:** Interaction of wild type and mutant RBD with human receptors. **A,** Interaction of RBD with ACE2. Here, green color represents ACE2, yellow color for wild type RBD and red color for mutant RBD. **B,** Interaction of RBD with NRP1. Here, blue color represents NRP1, yellow color for wild type RBD, and red color for mutant RBD.

**Fig 4:** Interaction of drugs with mutant Main Protease of Omicron variant. Amino acids of the binding site was presented with blue color. Here, A, interaction of Nirmatrelvir B, interaction of MPro 13b, and C, interaction of Lopinavir with Main Protease of Omicron variants.

**Fig 5:** 2D diagram of drug-protein interactions. Here, A, represent interaction between Nirmatrelvir and Main Protease, B, represent interaction between MPro 13b and Main Protease, C, represent interaction between Lopinavir and Main Protease.

1 **Table Legends**

3 **Table 1:** Nucleotide variations in all notable variants

4 **Table 2:** Consecutive deletions and insertions in omicron variant

5 **Table 3:** Mutations in the proteins encoded by omicron variant

6 **Table 4:** Binding energy of interaction between wild-type and mutant RBD with human receptors

7 **Table 5:** Binding energy of promising drugs against main protease of omicron variant

**Table 1: Nucleotide variations in all notable variants**

| Variants Name | Type | Total Mutations | Insertions | Deletions | Non-Synonymous |
|---|---|---|---|---|---|
| South Africa B.1.1.529 | Omicron | 117 | 9 | 52 | 50 |
| UK B.1.17 | Alpha | 58 | 0 | 19 | 23 |
| UK B.1.1.7+E484K | Alpha | 55 | 0 | 19 | 20 |
| UK B.1.1.7+L452R | Alpha | 52 | 0 | 19 | 20 |
| UK B.1.1.7+S494P | Alpha | 210 | 0 | 173 | 23 |
| South Africa B.1.351 | Beta | 48 | 0 | 18 | 11 |
| Netherlands B.1.351+E516Q | Beta | 59 | 0 | 28 | 23 |
| Canada B.1.351 | Beta | 42 | 0 | 18 | 20 |
| India B.1.617.2 | Delta | 35 | 0 | 0 | 22 |
| USA B.1.427 | Epsilon | 32 | 0 | 0 | 21 |
| Nigeria B.1.525 | Eta | 54 | 0 | 24 | 20 |
| Brazil P1 | Gamma | 38 | 0 | 0 | 12 |
| USA B.1.526 | Iota | 32 | 0 | 10 | 15 |
| USA B.1.526.1 | Iota | 38 | 0 | 13 | 21 |
| USA B.1.526.2 | Iota | 77 | 0 | 48 | 6 |
| India B.1.617.1 | Kappa | 36 | 0 | 0 | 24 |
| India B.1.617.3 | Kappa | 27 | 0 | 0 | 22 |
| Germany B.1.621 | Mu | 29 | 0 | 0 | 35 |
| France B.1.616 | Other | 69 | 0 | 32 | 25 |
| The Philippines P.3 | Other | 46 | 0 | 18 | 18 |
| Egypt C.36+L452R | Other | 37 | 0 | 6 | 25 |
| Russia AT.1 | Other | 144 | 12 | 95 | 15 |
| Switzerland B.1.1.318 | Other | 61 | 0 | 30 | 24 |
| UK A.23.1 | Other | 23 | 0 | 0 | 15 |
| Angola C.16 | Other | 29 | 0 | 0 | 18 |
| Belgium A.28 | Other | 33 | 0 | 6 | 16 |
| Italy A.27 | Other | 55 | 12 | 6 | 19 |
| Germany B.1.620 | Other | 67 | 0 | 18 | 15 |
| Argentina C.37 | Other | 42 | 0 | 9 | 14 |
| Belgium B.1.214.2 | Other | 69 | 9 | 32 | 19 |
| Brazil P.2 | Zeta | 46 | 0 | 19 | 13 |

**Table 2: Consecutive deletions and insertions in omicron variant**

| Mutation Types | Consecutive Base | Position | Effected Protein |
|---|---|---|---|
| Deletions | 3 | 6513 - 6515 | Papain like protease Nsp3 |
| Deletions | 9 | 11288 -11296 | Nsp6 |
| Deletions | 6 | 21765 - 21770 | Spike |
| Deletions | 9 | 21987 - 21995 | Spike |
| Deletions | 3 | 22194 - 22196 | Spike |
| Insertion | 9 | 22205 - 22213 | Spike |
| Deletions | 9 | 28395 - 28403 | Nucleocapsid Phosphoprotein |

**Table 3: Mutations in the proteins encoded by omicron variant**

| Protein Name | Mutations |
|---|---|
| Polyprotein ab (Papain-like Protease Nsp3) | Del:S2083; L2084I; A2710T |
| Polyprotein ab (Nsp4) | T3255I |
| Polyprotein ab (3C like Protease Nsp5) | P3395H |
| Polyprotein ab (Nsp6) | Del:L3674; Del:S3675; Del;G3676; I3758V |
| Polyprotein ab (RNA dependent RNA polymerase Nsp12) | P4715L |
| Polyprotein ab (Proofreading exoribonuclease nsp14) | I5967V |
| Spike | A67V; Del:H69; Del:V70; T95I; G142D; Del:V143; Del:Y144; N211I; L212V; In:213RE; V213P; R214E; G339D; S371L; S373P; S375F; K417N; G445S; S477N; T478K; E484A; Q493R; G496S; Q498R; N501Y; Y505H; T547K; D614G; H655Y; N679K; P681H; D796Y; N856K; Q954H; N969K; L981F |
| Envelope Protein | T9I |
| Membrane Glycoprotein | D3G; Q19E; A63T |
| Nucleocapsid Phosphoprotein | P13L; Del:E31; Del:R32; Del:S33; R203K; G204R |

**Table 4: Binding energy of interaction between wild and mutant RBD with human receptors**

| Human Receptors | Wild (kcal/mol) | Omicron (kcal/mol) |
|---|---|---|
| ACE2 | -15.9 | -17.2 |
| NRP-1 | -27.2 | -22.9 |
| DPP4 | -12.4 | -12.3 |
| BSG | -6.1 | -8.5 |

**Table 5: Binding energy of promising drugs against main protease of omicron variant**

| Drug Name | Binding Energy (kcal/mol) | |
|---|---|---|
| | Omicron | Wild |
| Ivermectin | -11.8 | -11.8 |
| Lopinavir | -9.6 | -9.5 |
| MPro 13b | -8.4 | -8.1 |
| Boceprevir | -8.4 | -9.6 |
| Ritonvir | -8.3 | -8.5 |
| GC-373 | -7.9 | -7.9 |
| Nirmatrelvir (Paxlovid) | -7.8 | -7.7 |
| GC-376 | -7.8 | -8.8 |
| PF-00835231 | -7.5 | -7.8 |
| MPro N3 | -7 | -7 |

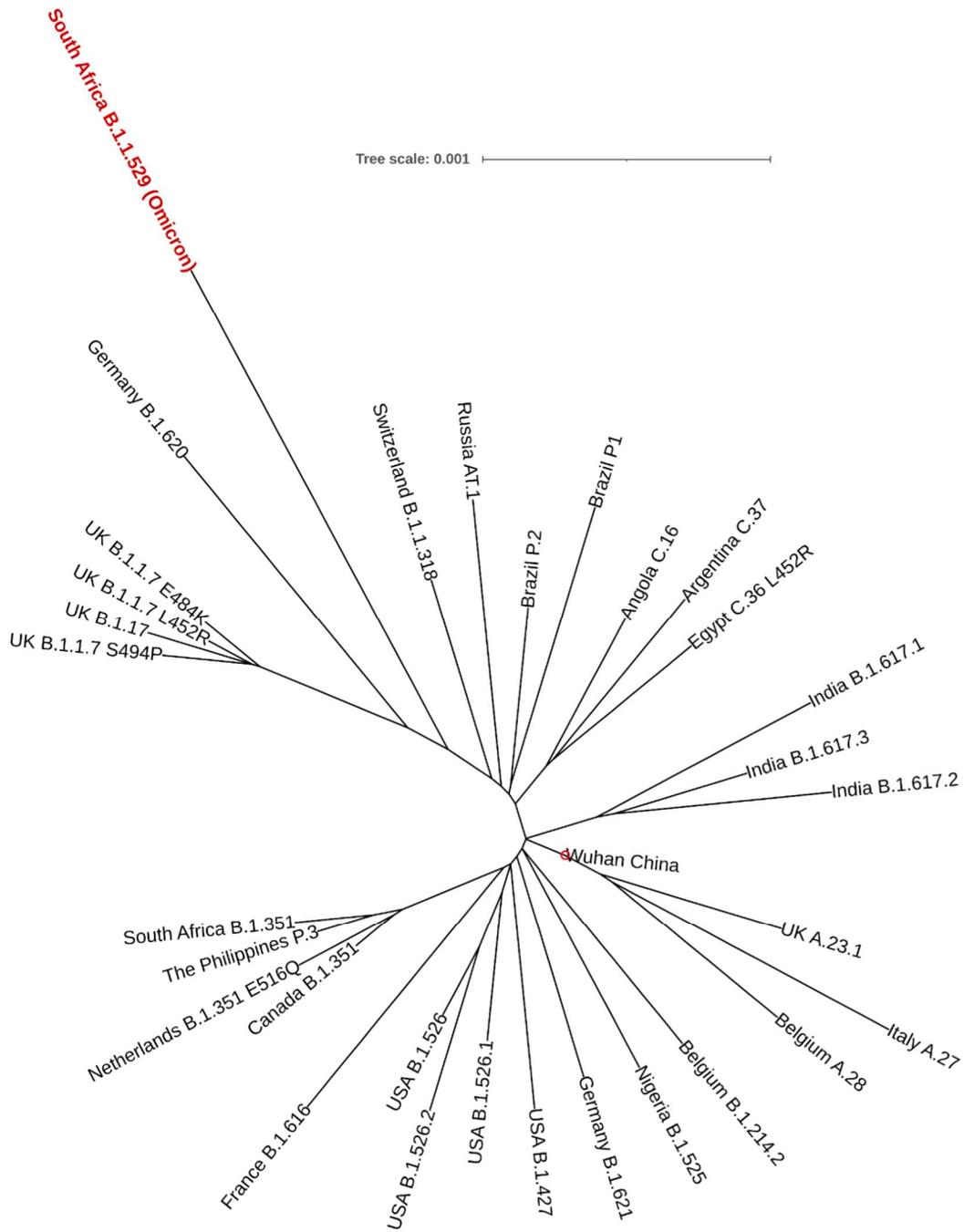

**Fig 1:** Maximum likelihood unrooted phylogenetic tree of all notable SARS-CoV-2 variants. South Africa B.1.1.529 (Omicron) was very close to Germany B.1.620. All the variants were centered with the Wuhan strain (red circle). A scale indicates genetic variation, defined as the number of substitutions per nucleotide site.

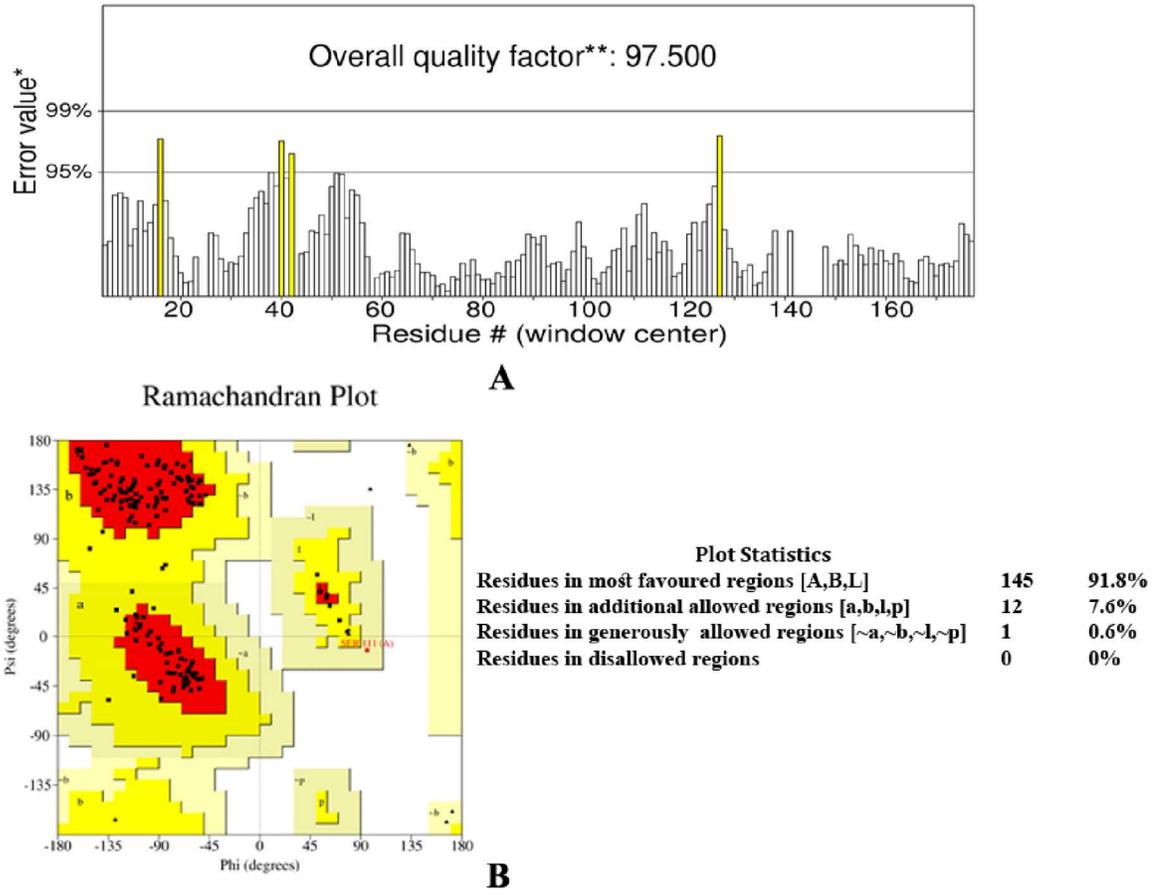

**Fig 2:** Validation of the 3D model. A, ERRAT validation assessment. B, Ramachandran Plot by PROCHECK.

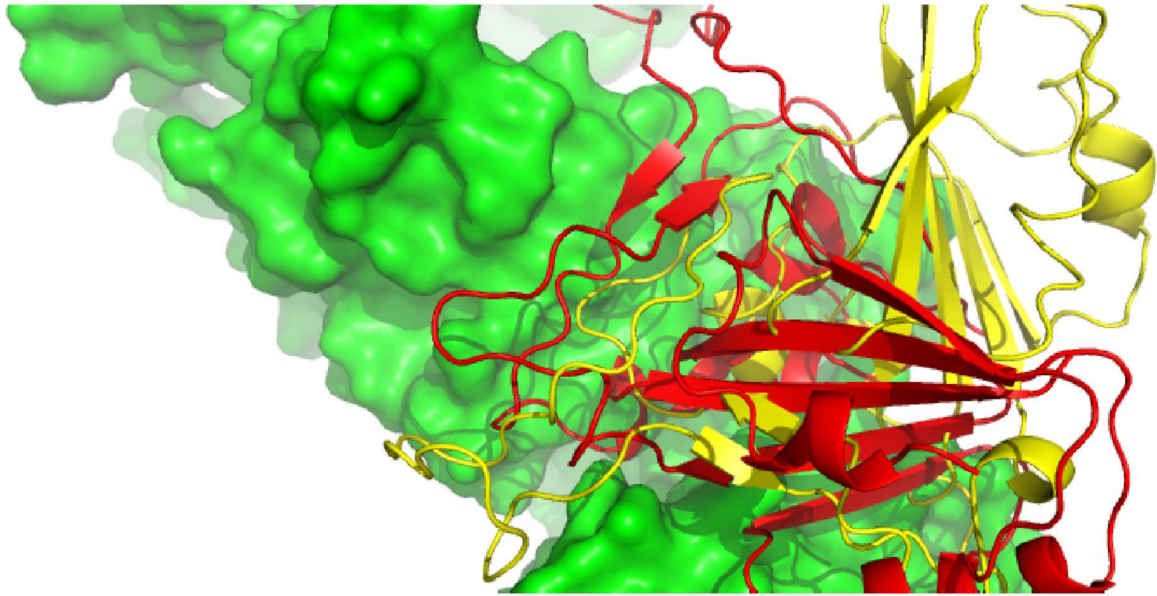

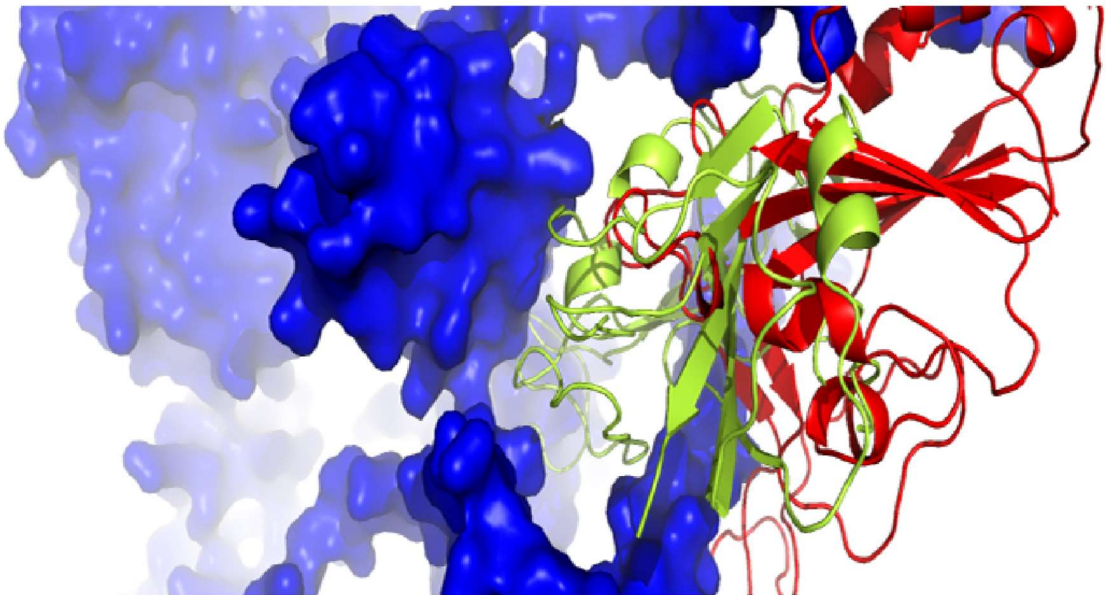

**Fig 3:** Interaction of wild type and mutant RBD with human receptors. A, Interaction of RBD with ACE2. Here, green color represents ACE2, yellow color for wild type RBD and red color for mutant RBD. B, Interaction of RBD with NRP1. Here, blue color represents NRP1, yellow color for wild type RBD, and red color for mutant RBD.

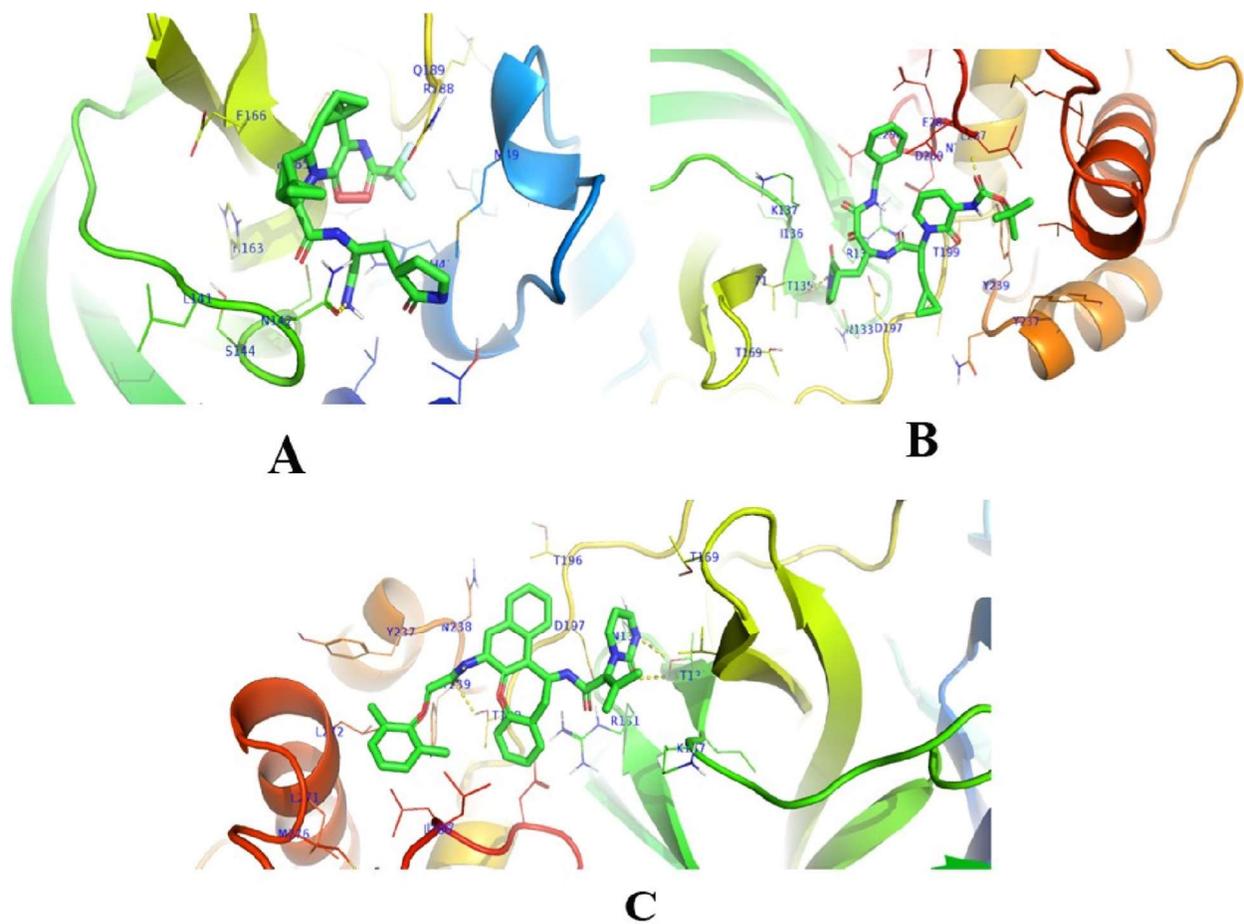

**Fig 4:** Interaction of drugs with mutant Main Protease of Omicron variant. Amino acids of the binding site was presented with blue color. Here, A, interaction of Nirmatrelvir B, interaction of MPro 13b, and C, interaction of Lopinavir with Main Protease of Omicron variants.

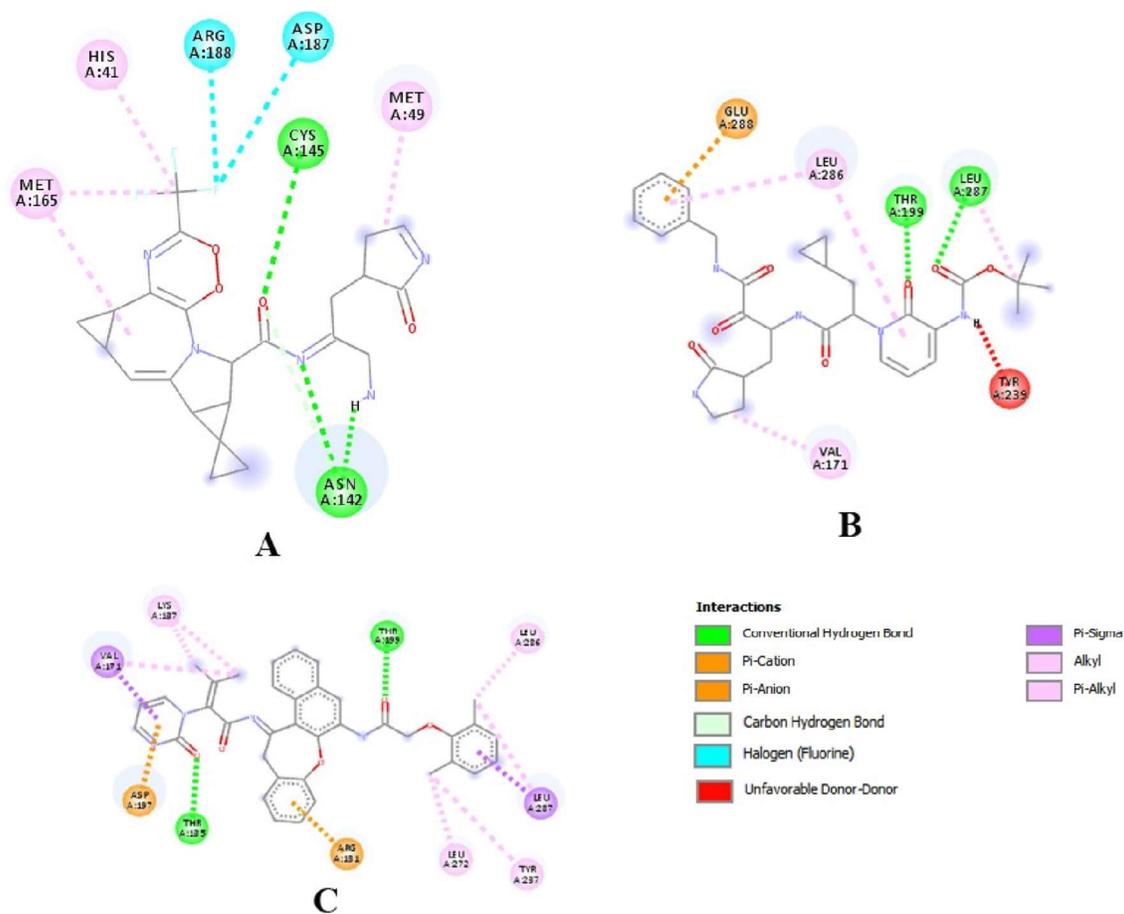

**Fig 5:** 2D diagram of drug-protein interactions. Here, A, represent interaction between Nirmatrelvir and Main Protease, B, represent interaction between MPro 13b and Main Protease, C, represent interaction between Lopinavir and Main P